\documentclass[%
 aapm,
 mph,%
 amsmath,amssymb,
 reprint,%
]{revtex4-2}

\usepackage{graphicx}
\usepackage{dcolumn}
\usepackage{bm}

\usepackage[mathlines]{lineno}
\modulolinenumbers[5]
\linenumbers\relax 

\usepackage{amsmath}
\usepackage{relsize}

\begin{document}

\nolinenumbers

\preprint{AAPM/123-QED}

\title[The relativistic Pythagorean three-body problem]{The relativistic Pythagorean three-body problem}

\author{Tjarda C. N. Boekholt}
\email{tjarda.boekholt@physics.ox.ac.uk}
\affiliation{ 
Rudolf Peierls Centre for Theoretical Physics, Clarendon Laboratory, Parks Road, OX1 3PU, Oxford, UK. 
}%

\author{Arend Moerman}
\author{Simon F. Portegies Zwart}
\thanks{All authors contributed equally to the work.}
\affiliation{%
Leiden Observatory, Leiden University, PO Box 9513, 2300 RA, Leiden, The Netherlands.
}%

\date{\today}

\begin{abstract}
We study the influence of relativity on the chaotic properties and 
dynamical outcomes of an unstable triple system; the Pythagorean three-body problem. 
To this end, we extend the \texttt{Brutus} N-body code to include Post-Newtonian pairwise terms up to 2.5 order, and the first order Taylor expansion to the Einstein-Infeld-Hoffmann equations of motion.  
The degree to which our system is relativistic depends on the scaling of the total
mass (the unit size was 1\,parsec).
Using the \texttt{Brutus} method of convergence, we test for time-reversibility in the conservative regime, and demonstrate that we are able to obtain definitive solutions to the relativistic three-body problem. It is also confirmed that the minimal required numerical accuracy for a successful time-reversibility test correlates with the amplification factor of an initial perturbation, as was found previously for the Newtonian case. 

When we take into account dissipative effects through gravitational wave emission, 
we find that the duration of the resonance, and the amount of 
exponential growth of small perturbations depend on the mass scaling.  For
a unit mass $\leq 10\,\text{M}_{\odot}$, the system behavior is
indistinguishable from Newton's equations of motion, and the
resonance always ends in a binary and one escaping body. For a mass
scaling up to $10^7 \text{M}_{\odot}$, relativity gradually becomes
more prominent, but the majority of the systems still dissolve in a
single body and an isolated binary.  The first mergers start to appear
for a mass of $\sim 10^5 \text{M}_{\odot}$, and between $10^7
\text{M}_{\odot}$ and $10^9$ $\text{M}_{\odot}$ all systems end
prematurely in a merger. These mergers are preceded by a
gravitational wave driven in-spiral.  For a mass scaling $\ge 10^9
\text{M}_{\odot}$, all systems result in a gravitational wave merger upon the first
close encounter. 
Relativistic three-body encounters thus provide an efficient pathway for resolving
the final parsec problem. 
The onset of mergers at the characteristic mass scale of $10^7\,M_\odot$
potentially leaves an imprint in the mass function of supermassive black holes. 
\end{abstract}

\maketitle

\section{Introduction}
\label{sec:intro}

The Newtonian three-body problem\cite{Newton1687} is one of the standard examples for illustrating chaos. This line of research dates back to Poincar\'e \cite{Poincare1892}, but more recent demonstrations have been made using computers. For instance, the series of papers started by \citet{Bahcall83} focuses on binary-single scattering experiments, in which a single body approaches a binary system with a certain impact parameter and relative velocity at infinity. They measure the cross section for the occurrence of a democratic resonance, e.g. a prolonged interaction between all three bodies with similar pair-wise gravitational forces. An example of the orbital chaos during such a resonance is given in their Fig. 3.  

A textbook example for a chaotic triple system is the Pythagorean problem, also called Burrau's problem \cite{Burreau1913}. In this case, the orbital chaos is not initiated by a binary-single encounter, but by three bodies performing a cold collapse, e.g. three bodies with masses
3, 4, and 5 are positioned at rest in a planar, right-angled triangle at
positions (1, 3), (-2, -1), and (1, -1), respectively.
In the subsequent evolution, the bodies fall towards the center of mass of the triple, after which the prolonged, chaotic interaction takes place. After a first close encounter between the three bodies, they experience multiple close encounters until the least massive body escapes, leaving the other two in a stable binary orbit. For a detailed discussion on the orbital evolution, we refer the reader to \citet{SP1967}.  

The Pythagorean problem exhibits exponential sensitivity to small changes in the initial condition\cite{AAOS93}. Over the lifetime of the interaction, small perturbations grow by about 9 orders of magnitude \cite{deJonghe1986, PB2018}, equivalent to about 20 e-folding time scales. This is one of the reasons why the 
Pythagorean three-body problem is excellently suitable for studying chaotic N-body systems numerically; the time scale on which perturbations grow is 
short compared to the system's lifetime. As a result, reprehensible \cite{PB2018} N-body codes have considerable difficulty in reaching a converged solution to such chaotic N-body problems, whereas for an arbitrary precise N-body code this can be achieved rather effortlessly \cite{BPV20}. 
Note that within this context, reprehensible or reprehensive refers to a solution to Newton's equations of motion for which the accumulation of numerical errors, 
and the system’s response, exceeds the exponential growth of the initial offset $\delta$ (see glossary by \citet{PB2018}).

The method adopted to measure the exponential sensitivity is based on that of Miller \cite{Miller64}, who considers an unperturbed N-body problem, and a perturbed one where a single phase space coordinate of a single body is offset in the n-th decimal place. By measuring the rate at which the 6N-dimensional phase space distance between the perturbed and the unperturbed solution grows (see their Eq. 2), we can establish if the growth is exponential, and determine the characteristic Lyapunov exponent. Alternatively, the total amplification factor, $A$, can be calculated as the ratio of the final to the initial magnitude of the perturbation, i.e. $A = \delta_{\rm{final}} / \delta_{\rm{init}}$ \cite{BPV20}. This exercise has been done for the Pythagorean problem\cite{PB2018}, as well as a more extensive ensemble of triple systems\cite{BPV20}, both under Newton's equations of motion. 

From an astrophysical point of view, the Newtonian three-body problem is accurate if other physical ingredients, such as tides, general relativity and stellar evolution,  have a negligible effect on the dynamics. Here, we relax one of these assumptions by reducing the physical scale and/or increasing the masses of the particles in such a way that Newton's approximation breaks down. In other words: if the velocities of the bodies start to approach the speed of light, general relativistic effects, such as precession and gravitational wave emission, have to be included in order to recover the correct  physical behavior. It was demonstrated, among others by \citet{Samsing2018}, that including these relativistic terms in the equations of motion in binary-single scattering experiments, has profound consequences for the merger rate of black holes, and consequentially on the occurrence of observable gravitational wave signals. In the same context of gravitational wave sources,  \citet{Rodri2018} demonstrated that including relativistic effects in the force calculations enhances the formation of binary black holes in star clusters. Dense stellar systems then become promising counterparts for ground based gravitational wave observatories (as was pioneered by \citet{2000ApJ...528L..17P}). 

Apart from affecting the dynamics of democratic triples, relativistic effects also change the dynamical behavior of hierarchical triples that are subject to von Zeipel-Lidov-Kozai resonant cycles \cite{2013ApJ...773..187N,2021MNRAS.500.3481H}. In the Solar System, Mercury's relativistic apsidal precession\cite{Einstein1915} quenches orbital resonances with Jupiter \cite{Laskar2009}. In each of these cases,  general relativity tends to stabilize the dynamical system; precession and energy dissipation through gravitational wave emission tend to drive the system towards an increasingly stable configuration. 

Relativistic effects in the Pythagorean problem have been studied by \citet{VMP1995}. They vary the masses of the bodies in units of solar mass, but fix the initial separations, which are considered to be in units of a parsec. Interestingly, they find a transition in the mass scale at around $10^7$ solar masses, below which the triple is ``escaper-dominated'', in the sense that one body escapes to infinity leaving behind an isolated binary. At higher masses the systems are ``merger-dominated'', e.g. one or two mergers occur as a result of gravitational wave mergers or head-on collisions. The merger-dominated regime is interesting with respect to the final parsec problem of merging supermassive black hole binaries\cite{MM03} as we discuss later.  

We aim to contribute to this field of research in the following ways: 1) we present a new version of the \texttt{Brutus} N-body code\cite{PB2014} including Post-Newtonian (PN) terms up to 2.5 order and 1~PN cross terms\cite{Einstein1938}, 2) we demonstrate that we can obtain numerically converged and time-reversible solutions to the conservative, relativistic three-body problem, and 3) we apply our new method to study the chaotic properties of the Pythagorean problem with varying mass scale as done previously by \citet{VMP1995}, but with numerically converged solutions, a higher resolution in initial condition space, and with extra information on the amplification factors, lifetimes and outcome space. In the next section we describe the \texttt{Brutus} N-body code with Post-Newtonian terms, in Sec.~\ref{sec:results1} we study the conservative, relativistic three-body problem and demonstrate time-reversibility, in Sec.~\ref{sec:results2} we study the influence of energy dissipation and demonstrate the transition from escaper to merger-dominated dynamics and the effect of relativity on the chaotic three-body problem. 

\section{Method}
\label{sec:method}

\begin{figure*}
\centering
\begin{tabular}{c}
\includegraphics[height=0.4\textwidth,width=0.9\textwidth]{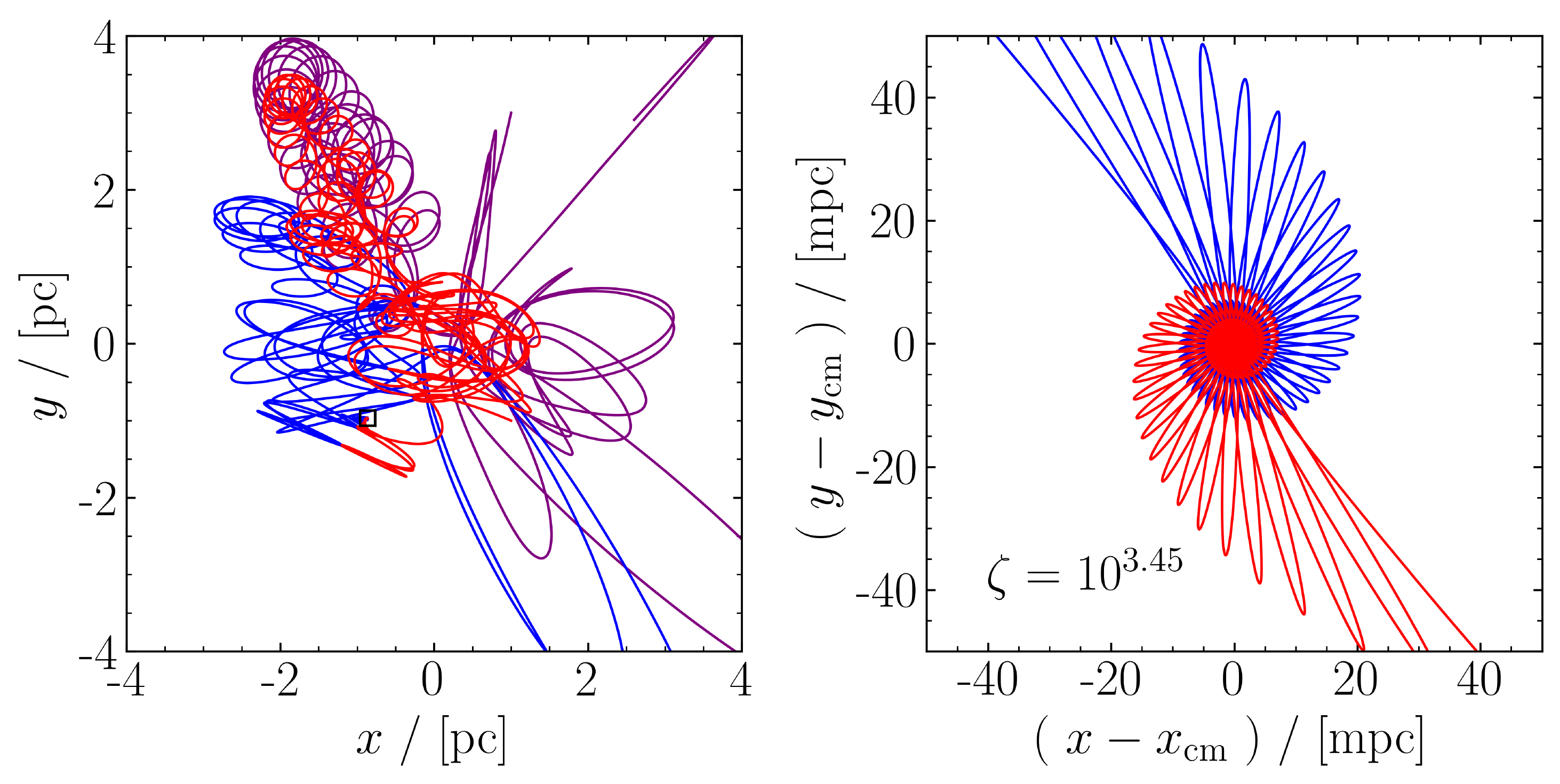} \\
\includegraphics[height=0.4\textwidth,width=0.9\textwidth]{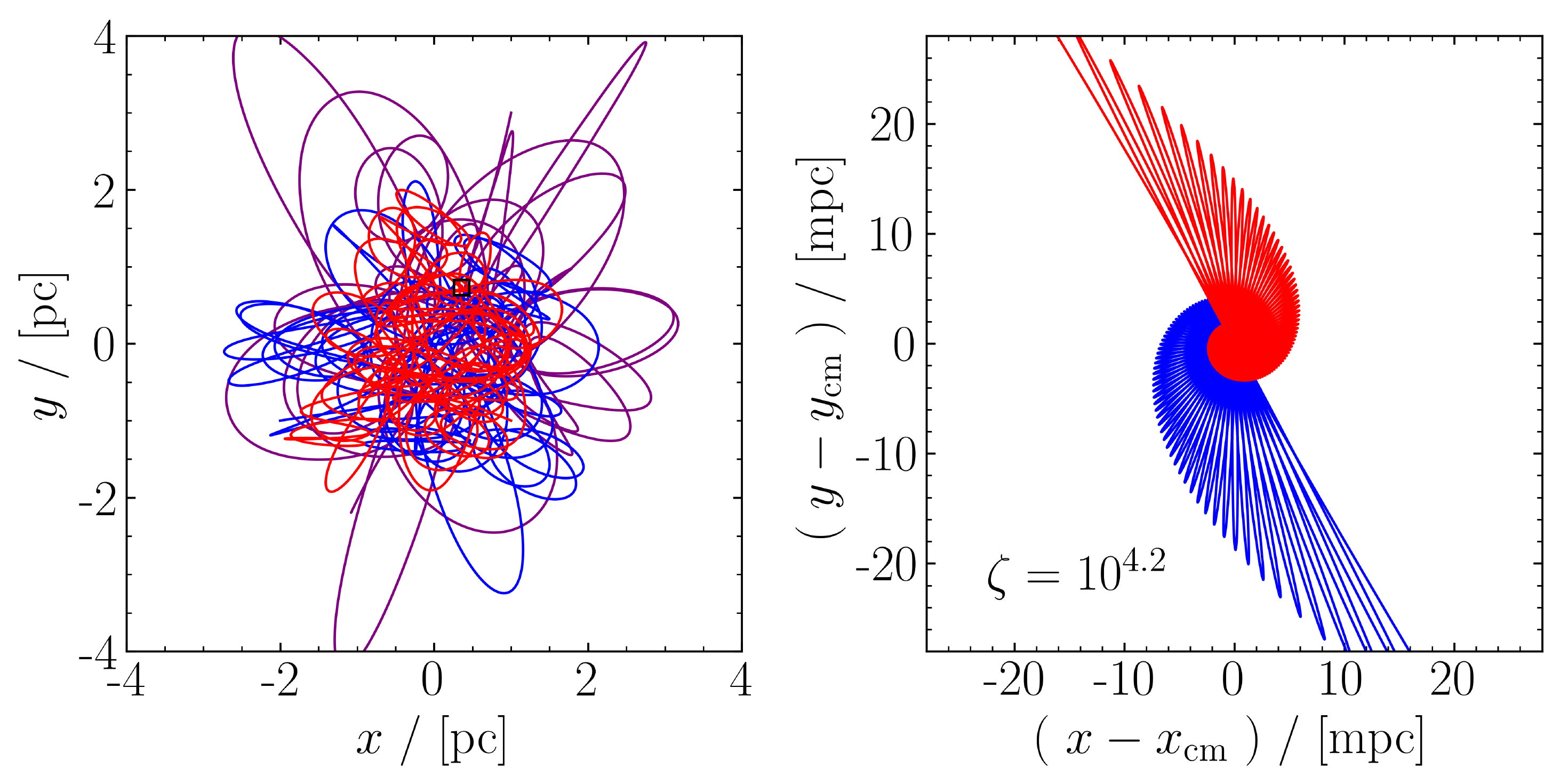} \\
\includegraphics[height=0.4\textwidth,width=0.9\textwidth]{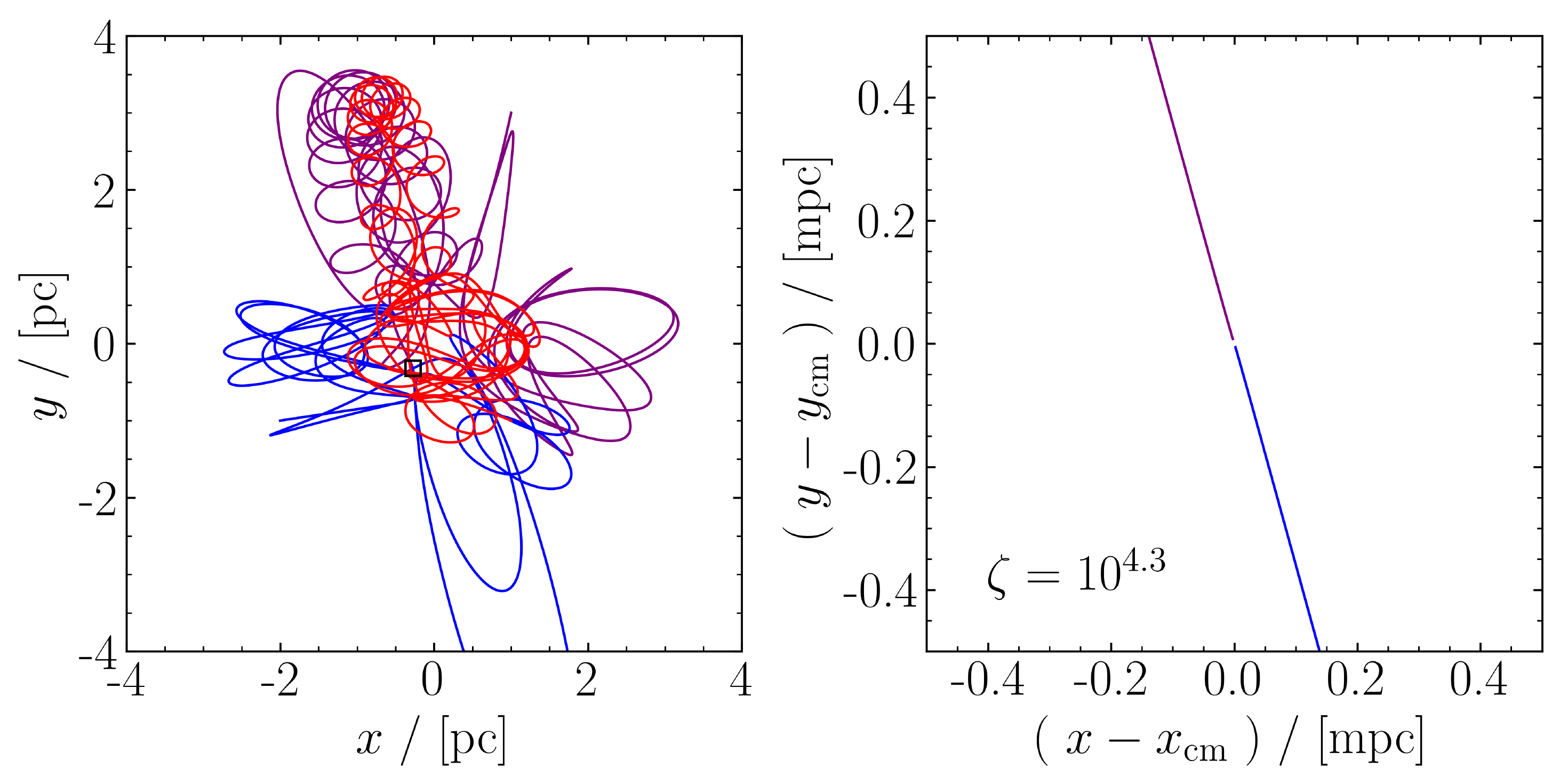} \\
\end{tabular}  
\caption{ Three solutions to the relativistic Pythagorean problem that end in a merger between two objects. Per row, we show the system's global chaotic behaviour in the left panel, and a zoom-in of the merger of two components in the right panel. The zoomed-in region is visualized by a small black box in the left panel, while the right panels are centralized at the center of mass of the merger progenitors. }
\label{fig:gallery}
\end{figure*}

The simulations are performed with the arbitrary-precision N-body code \texttt{Brutus}~\cite{PB2014}. This code implements the Bulirsch-Stoer~\cite{BS1964,Gragg1965} algorithm in order to control discretisation errors, and arbitrary-precision arithmetic in order to control round-off errors. The accumulation of numerical errors during a simulation cannot be avoided, but the aim of \texttt{Brutus} is to reduce the magnitude of the errors to below a threshold where it no longer affects the numerical results. We call this threshold the convergence limit $n$. For a simulation with a convergence limit of $n=3$, the first three decimal places in the final simulation results will remain the same when the integration time step is further reduced or the length of the mantissa is increased.

Associated with the two main ingredients mentioned above, are the two code parameters: 
the Bulirsch-Stoer tolerance, $\epsilon$, and the word-length, $L_w$, in bits. By systematically decreasing the value of $\epsilon$ and increasing $L_w$, numerically converged solutions are achieved, i.e. ``definitive solutions'' \cite{PB2018}. Conventional N-body codes are not able to obtain these definitive solutions, unless the simulation time is only a few Lyapunov time scales, or if the N-body configuration is not chaotic, such as in periodic braids \citep{moore_1993}. 

The underlying integration scheme of \texttt{Brutus} is the second-order, symmetric Verlet-Leapfrog method~\cite{Verlet1967}. This scheme works under the assumption that the underlying force field only depends on the mass and distances between bodies. However, the Post-Newtonian (PN) equations of motion explicitly depend on velocity. Therefore, the Verlet-Leapfrog scheme is replaced with the Auxiliary-Vector-Algorithm~\cite{HellstromMikkola2009}, or AVA for short (see Appendix \ref{app:ava}). This scheme is capable of handling velocity-dependent force fields using an explicit midpoint method to advance the velocities over one time step. The algorithm is symmetric under time reversal and reduces to the Verlet-Leapfrog when the force field only depends on the positions.
Note that the underlying integrator has to be time symmetric in order to facilitate the convergence in the Bulirsch-Stoer iteration. 

\begin{figure*}
\centering
\begin{tabular}{c}
\includegraphics[height=0.9\textwidth,width=0.9\textwidth]{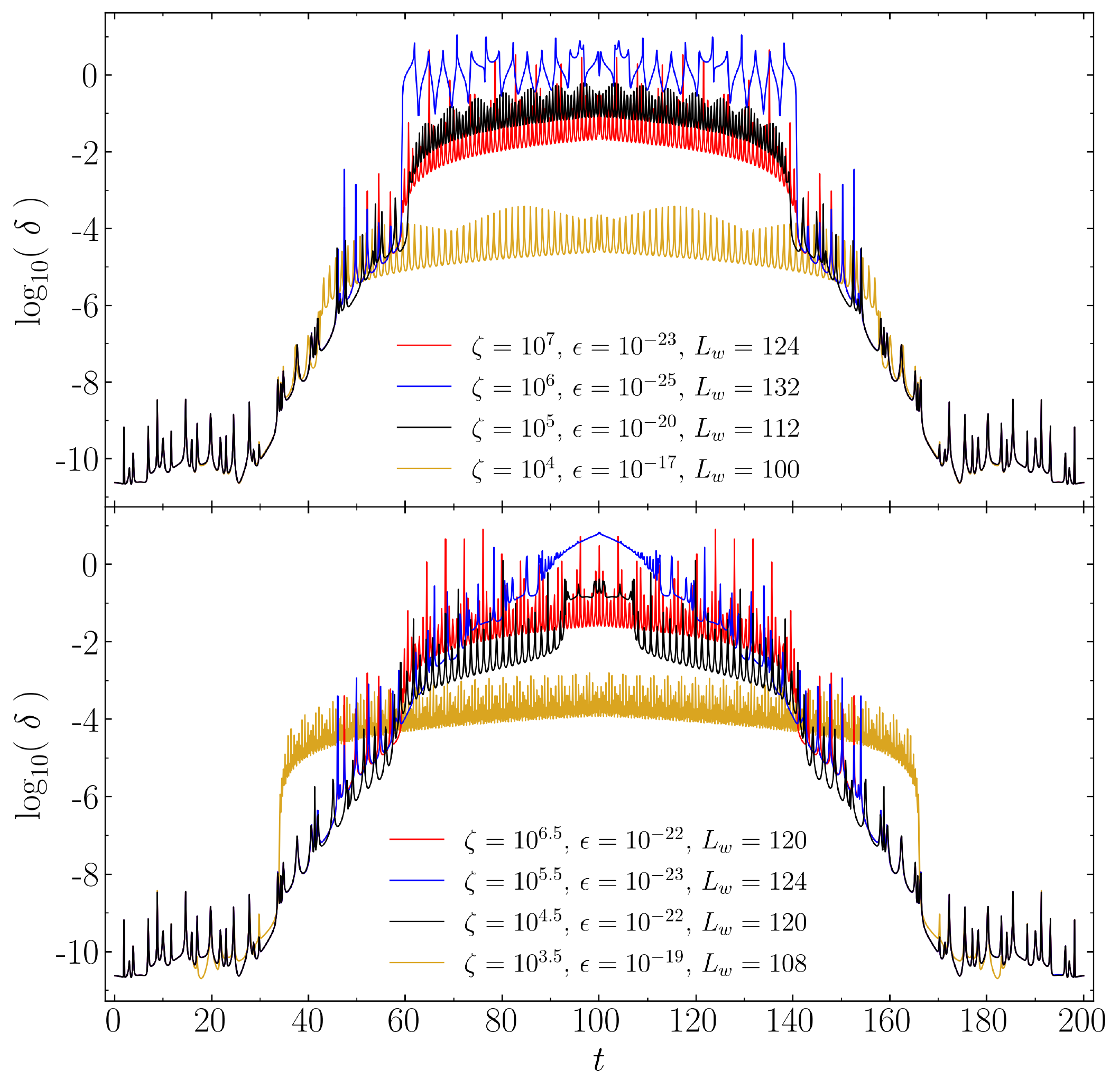} \\
\end{tabular}  
\caption{ Exponential growth of a small initial perturbation, $\delta$, as a function of time in the relativistic Pythagorean problem. Each curve was obtained with a different value of $\zeta$. A time reversible evolution, and thus a symmetric curve, was achieved for the specified values of $\epsilon$ and $L_w$. }
\label{fig:tophat}
\end{figure*}

The implemented PN acceleration terms include 1 to 2.5 PN pairwise and 1 PN cross terms\cite{Einstein1938} (see Appendix \ref{app:pn}). Using flags, we can control which terms to include in a simulation. By switching off the 2.5 PN term, for example, we are left with the conservative terms only. This allows us to check for energy conservation including PN corrections\cite{FutaItoh2007,Blanchet2014,Will2018}, and to determine the effect of relativity on time reversibility.

We adopt the initial conditions for the Pythagorean problem of three bodies with masses 3, 4 and 5 on the corners of a Pythagorean triangle with sides 3, 4 and 5 (see \citet{AAOS93} for a visualization of the triangle, and a study on the effect of small perturbations). The code further assumes a gravitational constant $G=1$ as is commonly adopted in N-body codes\cite{Heggie1986}. Rather than changing the values of the masses and separations, we instead vary the value of the dimensionless speed of light, $\zeta$. Hence, the specified value of $\zeta$ determines the ratio of velocity over speed of light, $v/c$. We convert to physical units by multiplying the positions by $f_r\,\rm{[pc]}$, with $f_r=1$, and the velocities by $f_v\,\rm{[km\,s^{-1}]}$, with $f_v = c / \zeta$, with $c$ the speed of light in physical units. The mass scale factor is given by $f_m = f_r f_v^2 / G$, with $G$ the gravitational constant in physical units.
Therefore, by systematically varying the value of $\zeta$, we are indirectly varying the physical mass scale in the Pythagorean problem, while keeping the size scale unit constant (at 1 parsec). 

In Fig.~\ref{fig:gallery} we show three solutions to the Pythagorean problem for $\log_{10}\,\zeta = 3.45, 4.2$ and $4.3$. These values correspond to $f_m = 2.63 \times 10^{6}$,  $8.32 \times 10^{4}$, and $5.25 \times 10^{4}$ solar masses, respectively. The left panels  demonstrate the orbital chaos in each solution, while the right panels show a zoom-in of the gravitational inspiral and precession (top two rows) and head-on collision for the most relativistic case (bottom row). 

\section{Time reversibility in chaotic relativistic triple systems}
\label{sec:results1}

Newton's laws of motion are symmetric with respect to the arrow of time. Time reversibility has therefore been used as a proxy for the accuracy of an N-body simulation. 
A successful time reversibility test for the Pythagorean problem was presented by \citet{PB2018}. Similar to \citet{Miller64}, they consider two realizations of the Pythagorean problem: the unperturbed one, and a perturbed one where the lightest body is slightly offset along the x-axis in the tenth decimal place (these are chosen arbitrarily). In the forward integration up to $t=100$, they measure an exponential divergence between the two trajectories in phase space. In the backward integration, exponential convergence to the initial size of the perturbation was only achieved for $\epsilon = 10^{-24}$, and $L_w = 128$. The required numerical accuracy and precision for achieving a time-reversible solution to a chaotic N-body problem is determined by the amplification factor of the initial perturbation\cite{BPV20}. Systems with a larger amplification factor require a higher numerical accuracy and precision.   

In a first experiment, we perform a time-reversibility test for the relativistic Pythagorean problem. The aim is to determine whether we can obtain time-reversible solutions to the relativistic Pythagorean problem, and to measure how relativistic terms affect the numerical accuracy needed to reach a converged solution. The experimental setup is the same as described above for the Newtonian case, but includes relativistic effects. 
Since time-reversibility can only be achieved in the absence of energy dissipation, we turn off the 2.5~PN term, and only include the 1~PN pairwise and cross terms, which are conservative.  

Each curve in Fig.~\ref{fig:tophat} represents the time evolution of the phase space distance between two neighboring trajectories, for a certain value of the dimensionless speed of light. The curves are symmetric around $t=100$, which corresponds to the moment we reversed the velocities. This confirms that the conservative, relativistic three-body problem is indeed time reversible. The shape of the curves varies with each value of $\zeta$. This can be understood by realizing that small differences in the force calculation will also grow exponentially due to the chaotic nature of the Pythagorean problem. Indeed, we see that the curves corresponding to $\zeta \ge 10^{4.5}$ start to deviate after about 40 time units from the Newtonian solution. For the strongest relativistic perturbations corresponding to $\zeta \le 10^4$, we observe that the deviation is visible even earlier, but that the maximum phase space separation is smaller by about three orders of magnitude. 

For a range of values for $\zeta$, we measure both the required numerical accuracy for convergence, $\epsilon$, and the amplification factor of the initial perturbation as was defined in Sec.~{\ref{sec:intro}}, e.g. the relative  height of the curves in Fig.~\ref{fig:tophat}. We measure an anti-correlation between the two quantities (see Appendix \ref{app:A_eps}), which was found previously in the Newtonian case\cite{BPV20}.  

\section{Relativistic triples as a solution to the final parsec problem}
\label{sec:results2}

In a second experiment, we aim to investigate the transition from escaper-dominated to merger-dominated dynamics, and the influence of relativity on the growth rate of small perturbations. To this end, we turn on all relativistic terms up to 2.5~PN, including the 1~PN cross terms. We integrate the Pythagorean problems until one of the following stopping conditions was satisfied: 1) one body was ejected and has escaped the system, or 2) there was a merger or collision between two bodies. The criteria for the first case is that the single body is: 1) at least 5 distance units away from the center of mass, 2) moving away from the center of mass, 3) has positive energy, and 4) satisfied all three previous constraints for at least 100 time units. The criterion for the second case is that the separation between two bodies has become less than the sum of their gravitational radii, given by $R_g = GM\,/\,c^2$. 

\begin{figure}
\centering
\begin{tabular}{c}
\includegraphics[height=0.81\textwidth,width=0.45\textwidth]{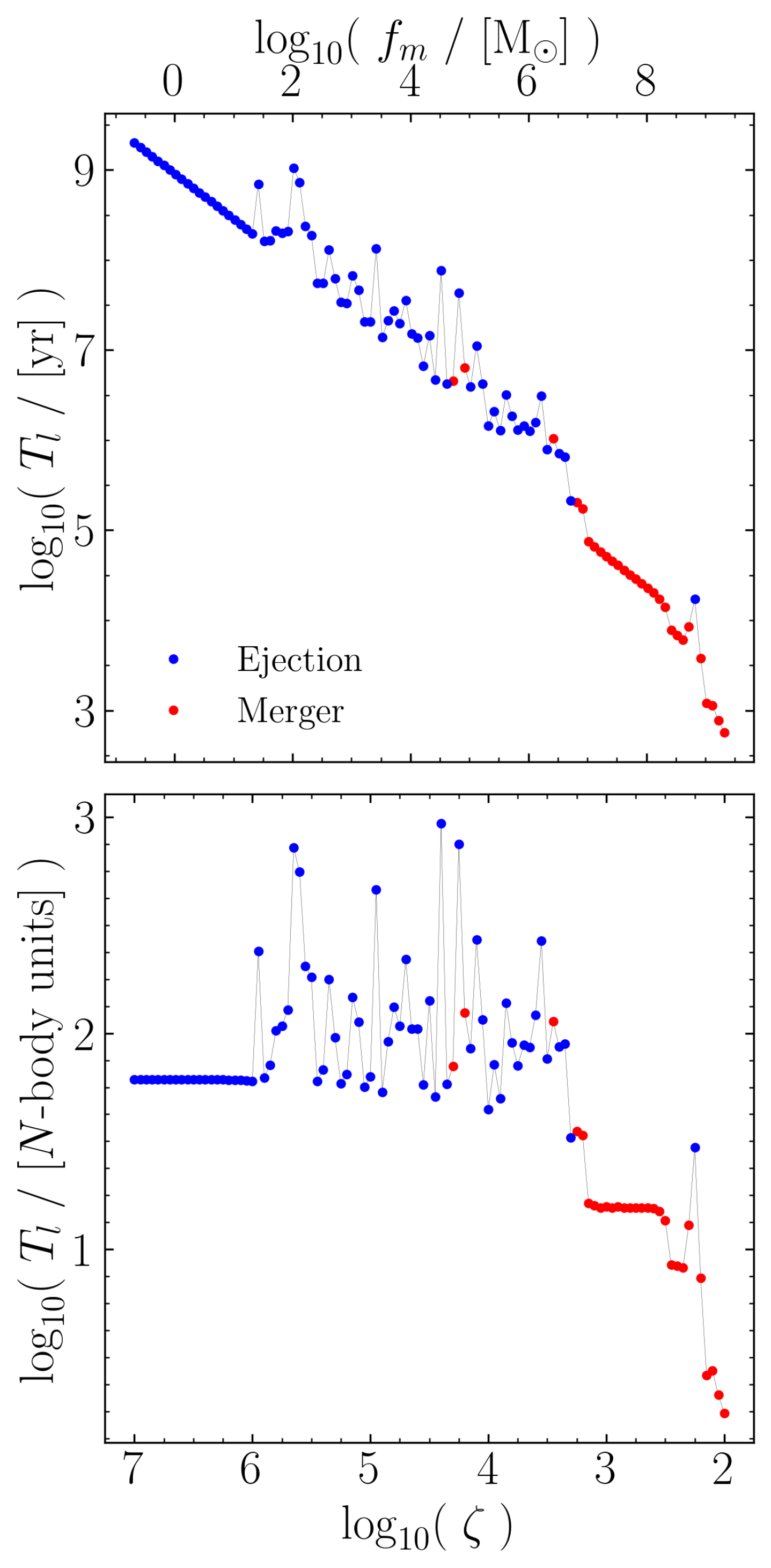} \\
\end{tabular}  
\caption{ Dependence of lifetime of the Pythagorean triple system on the strength of relativistic effects. The results in dimensionless quantities are given in the bottom panel, while the same results in physical units are given in the top panel. }
\label{fig:transition}
\end{figure}

In Fig.~\ref{fig:transition} we plot the lifetime as a function of dimensionless speed of light (bottom panel), or equivalently, the physical mass scale (top panel). In blue, we mark the systems which satisfy the escape stopping condition, while red marks the merger/collisional systems. We confirm the result obtained by \citet{VMP1995}, that there is a transition from escaper-dominated to merger-dominated dynamics at a mass scale of about $10^7$ solar mass. 

For values of $\zeta > 10^6$ (mass scale $< 10$ solar mass), the dimensionless lifetime is constant and consistent with the Newtonian solution. In this regime the relativistic perturbation remains negligible within the lifetime of the triples. For values of $\zeta$ ranging from $10^{3.5-6}$ (mass scale between $10^{1-7}$ solar mass), we observe a chaotic variation of lifetimes up to an order of magnitude. The combination of relativistic perturbations and their exponential magnification due to chaos, leads to a sensitive dependence in the lifetime. Note that in this chaotic blue sea of data points, there are also three red points indicating mergers. These are the three solutions visualized in Fig.~\ref{fig:gallery}. The chaotic motions of the three bodies can lead to such close encounters, that they result in gravitational wave captures and/or head-on collisions. This illustrates the importance of taking into account Post-Newtonian terms in the equations of motion for simulating the formation of gravitational wave sources. The merger time of the captured binary is shorter than the crossing time of the triple system. This is to be expected because the eccentricities of captured binaries tend to be high, as can be seen in Fig.~\ref{fig:gallery}. Finally, for values of $\zeta < 10^{3.5}$ (mass scales $>10^7$ solar mass), we observe a consistent evolution towards a merger. The time scale for change in orbital energy of the triple occurs on a time scale similar to the crossing time, resulting in short lifetimes and mergers.

\begin{figure}
\centering
\begin{tabular}{c}
\includegraphics[height=0.4\textwidth,width=0.5\textwidth]{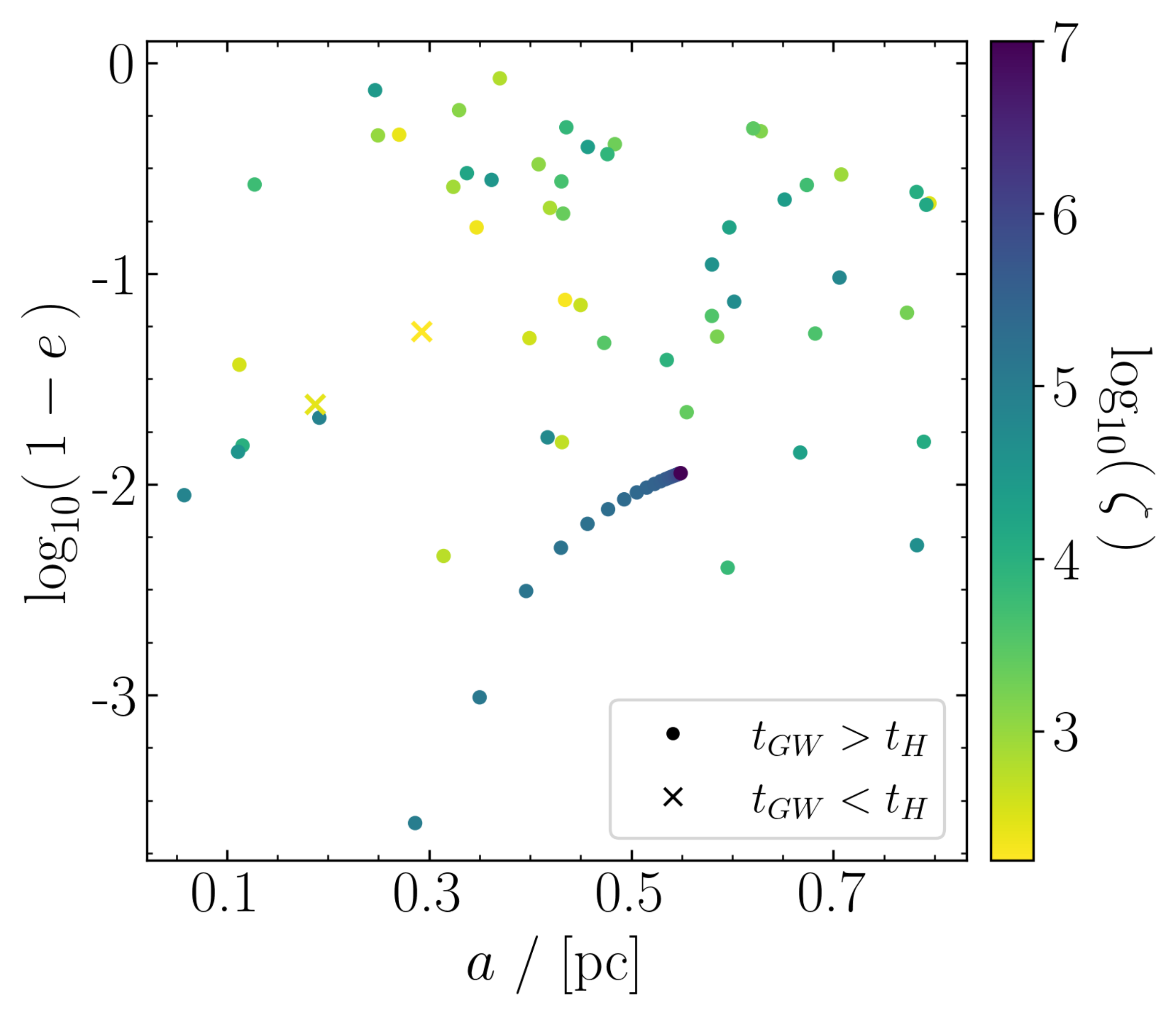} \\
\end{tabular}  
\caption{ Orbital elements of the ejected binary black holes. The dimensionless speed of light, $\zeta$, is a proxy for the total mass (decreasing $\zeta$ corresponds to increasing mass). The crosses denote binaries which merge within a Hubble time.   }
\label{fig:elem}
\end{figure}

In the red sea of dots in the merger-dominated regime in Fig.~\ref{fig:transition}, one blue dot can be found. This particular system results in an unbound binary-single pair. However, since this system is highly relativistic, the ejected binary still merges at a later time. To investigate this in more detail, we gathered all ejected binaries over the whole range of $\zeta$, and plotted their semi-major axes and eccentricities in Fig.~\ref{fig:elem}. Crosses are the ejected binaries, which merge within a Hubble time. We find only two crosses in the plot and they are both yellow, corresponding to the most relativistic systems. One of these systems corresponds to the lonely blue dot in the merger-dominated regime of Fig.~\ref{fig:transition}.     

\begin{figure}
\centering
\begin{tabular}{c}
\includegraphics[height=0.4\textwidth,width=0.5\textwidth]{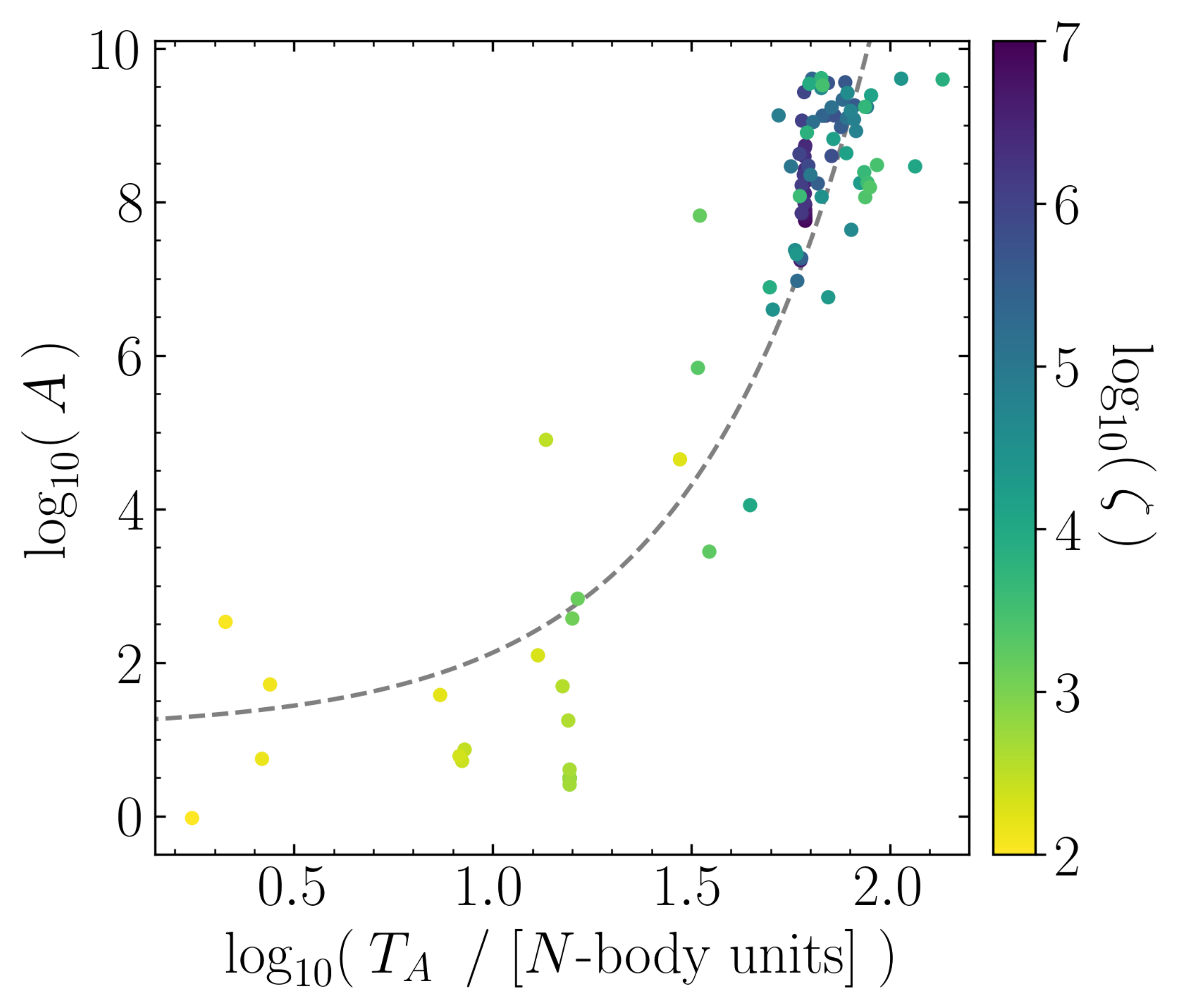} \\
\end{tabular}  
\caption{ Amplification factor as a function of amplification time. Here the amplification time, $T_A$, is defined to be the minimum between the lifetime, or the moment at which the phase space separation reached $\delta = 0.1$. We fit an analytical model with a constant Lyapunov time scale, $\log A = T_A / \tau_L + \gamma$, with $\tau_L = 4.3 \pm 0.2$ (see grey dashed curve). }
\label{fig:lyapunov}
\end{figure}

Finally, we consider the influence of relativity on the Lyapunov time scale of the Pythagorean problem. In our first experiment, we found that relativistic effects tend to stabilize the system in the sense that the amplification factors decreased for smaller values of $\zeta$. Since the amplification factor is given by $\log\,A = \lambda T$, with $\lambda$ the (time-averaged) Lyapunov exponent and $T$ the duration of the growth, we aim to determine whether the relativistic effects tend to change the Lyapunov exponent and/or the lifetime. For each triple in both our experiments, we measure the amplification time, $T_A$, taken to be the minimum of the lifetime or the time at which the phase space separation reached $\delta = 0.1$. Beyond this value, the growth is saturated, and the exponential growth cannot be measured reliably. We also measure the amplification factor of the initial perturbation at time $T_A$. The results are given in Fig.~\ref{fig:lyapunov}. We observe that triples in the Newtonian regime tend to lie in the top right, towards large amplification factors and relatively long lifetimes. As we gradually move the Pythagorean triangle into the relativistic regime, we find that the data approximately traces the dashed curve, which corresponds to a constant Lyapunov exponent: 

\begin{equation}
    \log_{10}\,A = \lambda_{10}T_A + \gamma,
\end{equation}

\noindent with $\lambda_{10} = 0.101 \pm 0.005$ and $\gamma = 1.1 \pm  0.3$. The corresponding Lyapunov exponent in base $e$ is $\lambda = 0.23 \pm 0.01$, while the Lyapunov time scale is the inverse, $\tau_L = 4.3 \pm 0.2$. 
We therefore conclude that relativistic effects mostly affect the lifetime of triple systems, rather than the Lyapunov time scale. The shorter lifetime results in less time for perturbations to grow, leading to smaller amplification factors. 

\section{Conclusions}
\label{sec:conclusions}

We present the new version of the arbitrary-precision N-body code \texttt{Brutus}\cite{PB2014}, which includes Post-Newtonian (PN) pairwise terms up to 2.5 order and 1~PN cross terms\cite{Will2018}. We demonstrate that this code is able to obtain definitive\cite{PB2018} solutions to the relativistic Pythagorean\cite{Burreau1913} three-body problem. 
By systematically increasing the mass scale in units of solar mass, but fixing the size scale to a parsec, we detect a characteristic mass scale of $10^7$ solar mass, confirming the result from \citet{VMP1995}. Below this value, relativistic perturbations grow exponentially, resulting in a sensitive dependence of the lifetime. The main  outcome is still an unbound binary-single configuration, but gravitational wave captures and head-on collisions start to occur at a mass scale of $10^5$ solar masses. Above the characteristic mass scale, a qualitatively different behavior is observed as the main outcome is the merger of two bodies due to a gravitational wave capture and subsequent in-spiral. The lifetimes also decrease driven by the dissipation of orbital energy, but the Lyapunov time scale remains unchanged to first order.  

The transitive mass scale of $10^7$ solar masses potentially plays an important role in solving the final parsec problem\cite{MM03, Gualan17} and the growth of supermassive black holes (SMBHs). 
Due to the hierarchical nature of galaxy mergers, triple SMBH systems might not be uncommon~\cite{Valtonen96,HL07, RPHOS18}. If the masses of the SMBHs fall below the characteristic mass scale, then the outcome of the triple interaction will most likely be a dynamical ejection. Depending on the ejection speed, the SMBH might escape the galaxy and terminate its growth, or fall back into the galactic nucleus through dynamical friction\cite{Chandra43I, Chandra43II, Chandra43III}, and engage in subsequent dynamical encounters\cite{MakFun04}. Alternatively, if the SMBHs are of the same order as the characteristic mass scale or above, mergers due to relativistic, three-body encounters become highly effective. The rate of growth due to the triple channel is then determined by the mass function of black hole seeds, and the merger rate of galaxies. The fact that below the characteristic mass scale other growth processes, such as gas accretion and runaway collisions with stars, dominate the growth of SMBHs, while above the characteristic mass scale, relativistic three-body encounters potentially dominate the growth through mergers, implies a potential imprint of the characteristic mass scale in the SMBH mass function\cite{Kelly2012}. 
This motivates further study into the growth rate of SMBHs through relativistic, three-body encounters, which take into account galaxy merger rates, initial mass functions for seed black holes, and other growth mechanisms, such as gas accretion.  

\begin{acknowledgments}
We thank Seppo Mikkola for suggesting the AVA integration method. 
This project was supported by funds from the European Research Council (ERC) under the European Union’s Horizon 2020 research and innovation program under grant agreement No 638435 (GalNUC). 
\end{acknowledgments}

\appendix

\section{Post-Newtonian equations of motion}\label{app:pn}

The Post-Newtonian (PN) accelerations implemented in the \texttt{Brutus} N-body code are taken from \citet{Blanchet2014,FutaItoh2007}, and the 1~PN cross terms from \citet{Will2018}. Here, we give the expressions up to 2.5~PN order (see Eq.~\ref{equation:PNeom}). All factors of the gravitational constant, $G$, are omitted, because we set $G=1$ throughout this paper. If one wants to include them, each mass $m$ should be multiplied by $G$. The body for which the acceleration is calculated is denoted by the subscript $i$, and $j$ ranges over all other bodies. The subscript $k$ denotes the summation index for the cross terms. This index ranges over all bodies unequal to $i$ or $j$.

\begin{widetext}
\begin{equation} \label{equation:PNeom}
\begin{split}
     \boldsymbol{a_{i}} = &-\mathlarger{\sum_{j\neq{i}}}\frac{m_{j}\boldsymbol{n}_{ij}}{r_{ij}^2} \\
     &+\frac{1}{c^2}\mathlarger{\sum_{j\neq{i}}}\frac{m_{j}\boldsymbol{n}_{ij}}{r_{ij}^2} \left[ 4\frac{m_j}{r_{ij}}+5\frac{m_i}{r_{ij}}+\mathlarger{\sum_{k\neq{i,j}}}\left(\frac{m_k}{r_{jk}}+4\frac{m_k}{r_{ik}}
    -\frac{m_{k}r_{ij}}{2r_{jk}^2}(\boldsymbol{n}_{ij}\cdot\boldsymbol{n}_{jk})\right)
    -v_{i}^2\right. \\
     &\quad\quad\left.+4\boldsymbol{v}_{i}\cdot\boldsymbol{v}_{j}-2v_{j}^2+\frac{3}{2}(\boldsymbol{v}_{j}\cdot\boldsymbol{n}_{ij})^2\right]-\frac{7}{2c^2}\mathlarger{\sum_{j\neq{i}}}\frac{m_j}{r_{ij}}\mathlarger{\sum_{k\neq{i,j}}}\frac{m_{k}\boldsymbol{n}_{jk}}{r_{jk}^2} \\
     &\quad\quad+\frac{1}{c^2}\mathlarger{\sum_{j\neq{i}}}\frac{m_j}{r_{ij}^2}\boldsymbol{n}_{ij}\cdot(4\boldsymbol{v}_{i}-3\boldsymbol{v}_j)(\boldsymbol{v}_i-\boldsymbol{v}_j) \\ 
     &+\frac{1}{c^4}\mathlarger{\sum_{j\neq{i}}}\frac{m_j\boldsymbol{n}_{ij}}{r_{ij}^2}\left[-2v_j^4+4v_j^2(\boldsymbol{v}_i\cdot\boldsymbol{v}_j)-2(\boldsymbol{v}_i\cdot\boldsymbol{v}_j)^2+\frac{3}{2}v_i^2(\boldsymbol{n}_{ij}\cdot\boldsymbol{v}_j)^2+\frac{9}{2}v_j^2(\boldsymbol{n}_{ij}\cdot\boldsymbol{v_j})^2\right. \\   
     &\quad\quad\left.-6(\boldsymbol{v}_i\cdot\boldsymbol{v}_j)(\boldsymbol{n}_{ij}\cdot\boldsymbol{v}_j)^2-\frac{15}{8}(\boldsymbol{n}_{ij}\cdot\boldsymbol{v}_j)^4-\frac{57}{4}\frac{m_i^2}{r_{ij}^2}-9\frac{m_j^2}{r_{ij}^2}-\frac{69}{2}\frac{m_im_j}{r_{ij}^2}\right. \\   
     &\quad\quad\left.+\frac{m_i}{r_{ij}}\left(-\frac{15}{4}v_i^2+\frac{5}{4}v_j^2-\frac{5}{2}(\boldsymbol{v}_i\cdot\boldsymbol{v}_j)+\frac{39}{2}(\boldsymbol{n}_{ij}\cdot\boldsymbol{v}_i)^2-39(\boldsymbol{n}_{ij}\cdot\boldsymbol{v}_i)(\boldsymbol{n}_{ij}\cdot\boldsymbol{v}_j)\right.\right. \\   
     &\quad\quad\left.\left.+\frac{17}{2}(\boldsymbol{n}_{ij}\cdot\boldsymbol{v}_j)^2 \right)\right. \\   
     &\quad\quad\left.+\frac{m_j}{r_{ij}}(4v_j^2-8(\boldsymbol{v}_i\cdot\boldsymbol{v}_j)+2(\boldsymbol{n}_{ij}\cdot\boldsymbol{v}_i)^2-4(\boldsymbol{n}_{ij}\cdot\boldsymbol{v}_i)(\boldsymbol{n}_{ij}\cdot\boldsymbol{v}_j)
    -6(\boldsymbol{n}_{ij}\cdot\boldsymbol{v}_j)^2)\right] \\   
     &+\frac{1}{c^4}\mathlarger{\sum_{j\neq{i}}}\frac{m_j\boldsymbol{v}_{ij}}{r_{ij}^2}\left[\frac{m_i}{r_{ij}}\left(\frac{55}{4}(\boldsymbol{n}_{ij}\cdot\boldsymbol{v}_j)-\frac{63}{4}(\boldsymbol{n}_{ij}\cdot\boldsymbol{v}_i)\right)-2\frac{m_j}{r_{ij}}((\boldsymbol{n}_{ij}\cdot\boldsymbol{v}_i)+(\boldsymbol{n}_{ij}\cdot\boldsymbol{v}_j))\right. \\   
     &\quad\quad\left.+v_i^2(\boldsymbol{n}_{ij}\cdot\boldsymbol{v}_j)+4v_j^2(\boldsymbol{n}_{ij}\cdot\boldsymbol{v}_i)-5v_j^2(\boldsymbol{n}_{ij}\cdot\boldsymbol{v}_j)-4(\boldsymbol{v}_i\cdot\boldsymbol{v}_j)(\boldsymbol{n}_{ij}\cdot\boldsymbol{v}_{ij})\right. \\   
     &\quad\quad\left.-6(\boldsymbol{n}_{ij}\cdot\boldsymbol{v}_i)(\boldsymbol{n}_{ij}\cdot\boldsymbol{v}_j)^2+\frac{9}{2}(\boldsymbol{n}_{ij}\cdot\boldsymbol{v}_j)^3\right] \\   
     &+\frac{1}{c^5}\mathlarger{\sum_{j\neq{i}}}\frac{4m_im_j}{5r_{ij}^3}\left[\boldsymbol{n}_{ij}(\boldsymbol{n}_{ij}\cdot\boldsymbol{v}_{ij})\left(\frac{52}{3}\frac{m_j}{r_{ij}}-6\frac{m_i}{r_{ij}}+3v_{ij}^2\right)+\boldsymbol{v}_{ij}\left(2\frac{m_i}{r_{ij}}-8\frac{m_j}{r_{ij}}-v_{ij}^2 \right) \right] \\   
     &+\mathcal{O}\left(\frac{1}{c^6}\right),
\end{split}
\end{equation}
\end{widetext}

\noindent where $\boldsymbol{a}_i$ is the acceleration vector of body $i$ and

\begin{equation}
    \begin{aligned}
        \label{equation:definitions}
        r_{ij}&=|\boldsymbol{r}_i-\boldsymbol{r}_j|, \\
        \boldsymbol{n}_{ij}&=\frac{\boldsymbol{r}_i-\boldsymbol{r}_j}{r_{ij}}, \\
        \boldsymbol{v}_{ij}&=\boldsymbol{v}_i-\boldsymbol{v}_j.
    \end{aligned}
\end{equation}

\noindent Here, $\boldsymbol{r}$ represents Cartesian positions and $\boldsymbol{v}$ Cartesian velocities.

\section{The Auxiliary-Vector-Algorithm}
\label{app:ava}

In order to extend the Newtonian \texttt{Brutus} N-body code to include Post-Newtonian terms, we have to extend the Verlet-Leapfrog integrator to include velocity-dependent forces, while preserving the time-symmetry. A natural extension is presented by \citet{HellstromMikkola2009} and is called the Auxiliary-Vector-Algorithm. This algorithm introduces an auxiliary velocity variable, which helps to make the algorithm explicit and time symmetric. For convenience, we reproduce the integration steps here: 

\begin{equation}
\label{equation:ava}
    \begin{aligned}
    \boldsymbol{r}_{i+\frac{1}{2}}&=\boldsymbol{r}_{i}+\frac{h}{2}\boldsymbol{v}_{i}, \\
    \boldsymbol{w}_{i+\frac{1}{2}}&=\boldsymbol{w}_{i}+\frac{h}{2}\boldsymbol{a}\big(\boldsymbol{r}_{i+\frac{1}{2}},\boldsymbol{v}_{i}\big), \\
    \boldsymbol{v}_{i+1}&=\boldsymbol{v}_{i}+h\boldsymbol{a}\big(\boldsymbol{r}_{i+\frac{1}{2}},\boldsymbol{w}_{i+\frac{1}{2}}\big), \\
    \boldsymbol{w}_{i+1}&=\boldsymbol{w}_{i+\frac{1}{2}}+\frac{h}{2}\boldsymbol{a}\big(\boldsymbol{r}_{i+\frac{1}{2}},\boldsymbol{v}_{i+1}\big), \\
    \boldsymbol{r}_{i+1}&=\boldsymbol{r}_{i+\frac{1}{2}}+\frac{h}{2}\boldsymbol{v}_{i+1}, \\
    \end{aligned}
\end{equation}

\noindent where $\boldsymbol{r}$ is the position, $\boldsymbol{v}$ the velocity and $\boldsymbol{a}$ the acceleration of the body which is being evolved and $h$ is the timestep. The auxiliary vector $\boldsymbol{w}$ is used to advance the real velocity vector $\boldsymbol{v}$ in time using a modified midpoint approach. 

\section{Initial configurations}
\label{sec:initcond}

The initial configurations used in this paper are identical to those used by \citet{PB2018}. The unperturbed initial condition is the standard Pythagorean problem, while in the perturbed version, we introduce an offset in the x-coordinate of the lightest body in the tenth decimal place. After integrating forward up to $t=100$ and then integrating backwards for the same amount of time (or reversing the sign of the velocities), we should in principle arrive at the initial condition again. In Table~\ref{tab:initref}, we give the final condition for a run with dimensionless speed of light $\zeta = 10^6$, at the precision used in the numerical integration.  

\begin{table*}
\caption{Initial and final conditions of the (un)perturbed reversibility test associated with $\zeta = 10^6$.}
\begin{center}
\begin{tabular}{ |l|l|l|l|l| } 
\hline
& Initial & Initial & Final   & Final \\
& unpert. & pert.   & unperturbed & perturbed \\
\hline
$x_3$  & \hspace{1mm}1.0 & \hspace{1mm}1.0+$10^{-10}$ & \hspace{1mm}0.999999999999877092833786140987105719279 & \hspace{1mm}1.000000000099912600957529245272663073222 \\
$y_3$  & \hspace{1mm}3.0 & \hspace{1mm}3.0 & \hspace{1mm}2.999999999999846567192885715279183049502 & \hspace{1mm}2.999999999999890883904268165911218676537 \\
$vx_3$ & \hspace{1mm}0.0 & \hspace{1mm}0.0 & \hspace{1mm}9.574369343984002518157266122431769280988$\times10^{-14}$ & \hspace{1mm}6.808960671291115536982119119626051608586$\times10^{-14}$ \\
$vy_3$ & \hspace{1mm}0.0 & \hspace{1mm}0.0 & -2.013464927983385791018810039102584729002$\times10^{-13}$ & -1.431906572143561770232050609564912599413$\times10^{-13}$ \\
$x_4$  & -2.0 & -2.0 & -2.000000000000105782093923622878780404431 & -2.000000000000075228565455684518296831068 \\
$y_4$  & -1.0 & -1.0 & -1.000000000000800560567546657459324553343 & -1.000000000000569330999227957938478115590 \\
$vx_4$ &  \hspace{1mm}0.0 &  \hspace{1mm}0.0 & -2.013464894027148639921316234636568914246$\times10^{-13}$ & -1.431906546964266790802223094781426070001$\times10^{-13}$ \\
$vy_4$ &  \hspace{1mm}0.0 &  \hspace{1mm}0.0 & -9.574369343931415002666762885031393616285$\times 10^{-14}$ & -6.808960671610479724054740650464272347989$\times10^{-14}$ \\
$x_5$  &  \hspace{1mm}1.0 &  \hspace{1mm}1.0 & \hspace{1mm}1.000000000000158369974864986257658181499 & \hspace{1mm}1.000000000000112627242267676309136378774 \\
$y_5$  & -1.0 & -1.0 & -0.999999999999267491861693451844884555278 & -0.999999999999479065543178069974885190030 \\
$vx_5$ &  \hspace{1mm}0.0 &  \hspace{1mm}0.0 & \hspace{1mm}1.036309754579876527859666306416566085918$\times10^{-13}$ & \hspace{1mm}7.369875972919536493593051989123011909700$\times10^{-14}$ \\
$vy_5$ &  \hspace{1mm}0.0 &  \hspace{1mm}0.0 & \hspace{1mm}1.974028504304201349271608972171993742073$\times10^{-13}$ & \hspace{1mm}1.403860797014731278786369987132584068207$\times10^{-13}$ \\
\hline
\end{tabular}
\label{tab:initref}
\end{center}
\end{table*}

\section{Anti-correlation between amplification factor and numerical accuracy}
\label{app:A_eps}

\begin{figure}
\centering
\begin{tabular}{c}
\includegraphics[height=0.4\textwidth,width=0.5\textwidth]{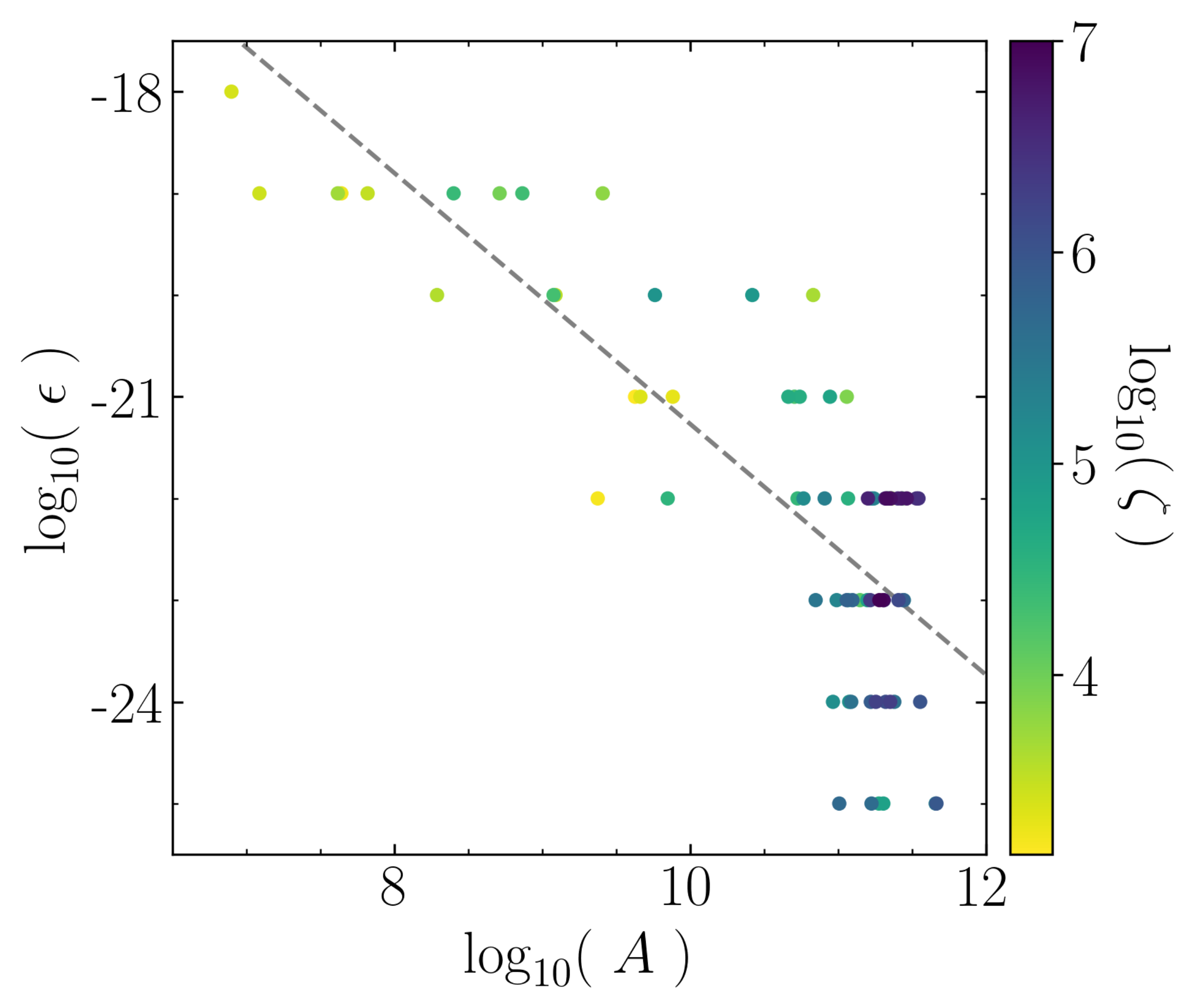} \\
\end{tabular}  
\caption{ Numerical accuracy required to achieve a time-reversible solution as a function of amplification factor. The linear fit has a slope of $-1.23 \pm 0.09$, which confirms the anti-correlation between numerical accuracy and amplification factor. Deeper into the relativistic regime (smaller values of $\zeta$), the triples show less chaotic behaviour (smaller amplification factors).  }
\label{fig:logA}
\end{figure}

For a given ensemble of triple systems, the irreversible fraction can be derived from the distribution of amplification factors\cite{BPV20}. We test this for the case of the relativistic Pythagorean problem. In Fig.~\ref{fig:logA} we plot the amplification factor as a function of the minimum required numerical accuracy for reaching a converged solution. A linear fit gives a slope of $-1.23 \pm 0.09$ which confirms the anti-correlation. The colour bar specifies the values of $\zeta$ and we observe a clear gradient in the sense that increasingly relativistic triples have smaller amplification factors. This is in line with the expectation that relativistic effects tend to reduce chaos in the dynamical system, see also \citet{PZ21}.  

\nocite{*}
\bibliography{cas-refs}

\end{document}